# Crystal Nucleation in an AlNiZr Metallic Liquid: Within and Beyond Classical Nucleation Theory

Fangzheng Chen,[1] Kian Cole Dahlberg[2], Zohar Nussinov[2,3] and K. F. Kelton[2,1,*]

1. Institute of Materials Science and Engineering, Washington University in St. Louis, St. Louis, Missouri 63130 USA
2. Department of Physics, Washington University in St. Louis, St. Louis, Missouri 63130 USA
3. Rudolf Peierls Centre for Theoretical Physics, University of Oxford, Oxford 0X1 3PU, United Kingdom
*. Corresponding author

**Abstract**

The Classical Nucleation Theory (CNT) has played a key role in crystal nucleation studies since the 19th century and has significantly advanced the understanding of nucleation. However, certain key assumptions of CNT, such as a compact and spherical nucleating cluster and the concept of individual diffusive jumps are questionable. The results of molecular dynamics (MD) studies of crystal nucleation in a $Al_{20}Ni_{60}Zr_{20}$ metallic liquid demonstrate that the nucleating cluster is neither spherical nor compact. The seeding method was employed to determine the critical cluster size and nucleation parameters from CNT, which were then compared to those derived from the Mean First Passage Time (MFPT) method. While the CNT-based nucleation rate aligns well with experimental data from similar metallic liquids, the MFPT rate differs significantly. Further, contrary to the assumption of individual jumps for atoms to join the nucleating cluster, a cooperative mechanism of attachment or detachment is observed. This is accompanied by synchronized changes in the local potential energy. Similar cooperative motion also appeared in a non-classical nucleation process, particularly during the coalescence of nuclei.

## 1. Introduction

Most first-order phase transitions are initiated by nucleation, where small regions with an order parameter that characterizes a new phase are stochastically formed. Understanding and controlling crystal nucleation in liquids are essential in many areas of chemistry, materials science and physics. Heterogeneous nucleation occurs at specific sites, while homogeneous nucleation, which is the focus of the studies discussed here, occurs randomly in space and time. Although it is difficult to make experimental studies of homogeneous nucleation due to reactions between the liquid and container, particularly for metallic liquids at high temperatures, containerless techniques have allowed some quantitative studies to be made.

Nucleation is commonly modeled within the framework of the Classical Nucleation Theory (CNT). A barrier to nucleation was first evident in the supercooling experiments of Fahrenheit, which showed that water could be kept in the liquid phase at temperatures well below its melting temperature for an extended period of time without crystallizing to ice. More than 150 years ago Gibbs developed a thermodynamic model for liquid nucleation in a gas that is based on this concept of a nucleation barrier. This model forms the basis for CNT. Gibbs assumed that the barrier arose from the energy that was required to create an interface between the nucleating cluster and the parent phase. He assumed that this interface is sharp and that the nucleating cluster is spherical and compact. His model leads to the concept of a critical size, $n^*$, for which the work of cluster formation, $W^*$, is a maximum. Gibbs argued that the nucleation rate is proportional to $\exp(-W^*/k_B T)$, where $k_B$ is Boltzmann's constant and $T$ is the temperature in absolute units. However, nucleation is also a kinetic process. The kinetic model embedded in CNT was proposed by Volmer and Weber, assuming that nucleating clusters shrink and grow by a series of bi-molecular processes with single molecules attaching or detaching at each step [1]. While CNT was originally developed to describe gas condensation, it was later extended by Turnbull and Fisher to describe crystal nucleation in a supercooled liquid [2], retaining the thermodynamic and kinetic assumptions of CNT. They also assumed that the kinetics of interfacial attachment were determined by a diffusive-type jump from the liquid onto the cluster, with a rate determined by the diffusion coefficient in the original phase.

Several studies have raised questions about the validity of many of the assumptions made in CNT. For example, experimental [3] and density functional theory (DFT) [4] studies indicate that the interface is not sharp. Further, DFT studies indicate that for small clusters the order parameter may not be representative of that of the bulk crystal [5]. A recently proposed analytical model suggests that cluster growth does not occur through individual atomic attachments, but rather through the cooperative attachment of multiple atoms [6, 7]. Molecular dynamics (MD) studies in several metallic liquid alloys suggest that this is also true for nucleation [8], with clusters growing or dissolving through the collective behavior of groups of 5 to 10 atoms. Further, recent in situ observations have shown that cluster growth can occur by the coalescence of nuclei, which is a non-classical mechanism [9, 10]. There has been little exploration of possible cooperative motion among the interface atoms during this coalescence. Is cooperative motion exclusive to nucleation? Does it feature in the coalescence of nuclei?

In this paper, these questions and the assumptions made in the development of the CNT are examined based on MD studies in the $Al_{20}Ni_{60}Zr_{20}$ metallic liquid. We will show that small nucleating clusters are neither spherical nor compact, and the order parameter decreases from the cluster center to the interface, in agreement with earlier work. Further, the order in the center of small clusters is considerably less than that of larger clusters that are more representative of the bulk crystal. The critical sizes and nucleation rates obtained from seed studies in the liquid reasonably agree with experimental results when CNT is assumed and when the kinetics are described in terms of the diffusion kinetics. However, the commonly used Mean First Passage Time (MFPT) method to obtain nucleation rates from MD simulations [11, 12] yields values that are orders of magnitude larger. New results for the cooperative attachment of clusters during nucleation and the cooperative rearrangement in the interface of two coalescing clusters are also presented.

## 2. Method of MD simulation

The MD simulations were made using the Large-scale Atomic/Molecular Massively Parallel Simulator (LAMMPS) in the Extreme Science and Engineering Discovery Environment [13]. An ensemble of the $Al_{20}Ni_{60}Zr_{20}$ metallic liquid containing 25,000 atoms was constructed by randomly locating 15,000 nickel, 5,000 aluminum, and 5,000 zirconium atoms. The sample was initially heated to 2,500K and maintained at this temperature for 2 ns. Subsequently, it was cooled to 800K at a rate of 10K/ps. As the temperature decreased, dump files with atomic information were generated at various temperatures ranging from 1250K to 900K. After reaching the target temperature in this range the sample was relaxed to study the nucleation and growth process using both a seeding method and spontaneous cluster generation (homogeneous nucleation). The atomic interactions were described using the Embedded Atom Potential (EAM) developed by Ward [14].

This potential has been validated in previous work from our group [8]. All simulations were made using the NPT (isobaric-isothermal) ensemble with periodic boundary conditions.

**3. Results and discussion**

3.1. Cluster Compactness and Interfacial Width

The $Al_{20}Ni_{60}Zr_{20}$ metallic liquid was held at temperatures ranging from 900K to 1050K for 2 to 10 ns to observe homogenous nucleation. The cluster/liquid interface was analyzed in terms of an order parameter equal to the dot product of the bond-orientational order, Q6 [15, 16],

$$\vec{q}_6(i) \cdot \vec{q}_6(j) = \sum_{m=-6}^{m=6} \tilde{q}_{6m}(i) \tilde{q}_{6m}(j)^* , \qquad (1)$$

where $\tilde{q}_{6m}$ is the normalized local orientational order parameter. The dot product indicates the similarity of the local environment for neighboring atoms $i$ and $j$. This parameter, termed the index of crystallinity (IC), demonstrated an efficacy for distinguishing between crystal and liquid atoms in our prior research [8]. The cutoff for the IC calculation was set at 7.5 Angstroms (Å). Using this cutoff, a value of 120 aligns well with the coordination numbers calculated from the radial pair distribution.

The geometry of the emerging cluster was analyzed by observing the cluster density distribution, visualized using OVITO [17]. This is depicted as two-dimensional scatter plots that show the number density of different clusters in fig. 1. As shown in fig. 1.a smaller clusters are irregularly shaped and do not have a distinct dense center, which is in contrast with larger clusters. As the size of the cluster increases the highest number density is centrally located and the number density decreases as the distance from the center increases. These observations align with predictions from DFT calculations [5] and experimental studies of colloidal crystallization [3]. However, even the larger clusters shown in fig. 1.b – 1.d exhibit a clear asymmetry, which challenges the compact spherical assumption made in the CNT.

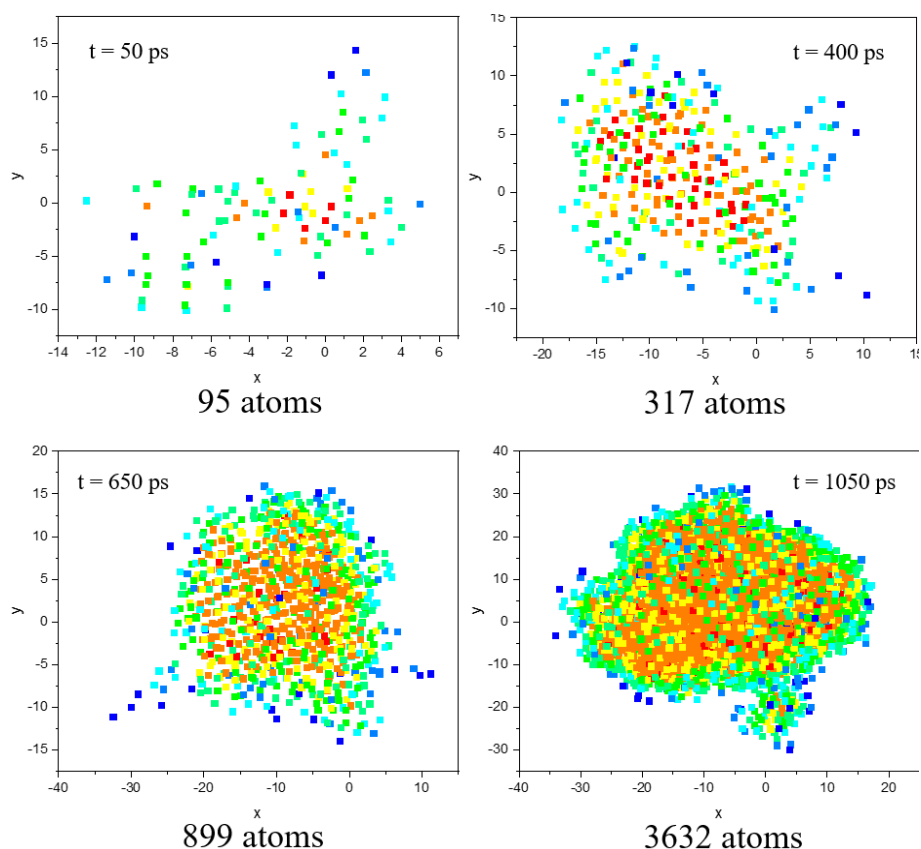

**Figure 1-** Growing cluster at 1050K projected onto the x-y plane. (a) A cluster containing 95 atoms; (b) a cluster containing 317 atoms; (c) a cluster containing 899 atoms; (d) a cluster containing 3632 atoms. Regions highlighted in red indicate the highest relative number density, whereas blue denotes areas of the lowest density. The axis is labeled with position coordinates that are measured in Angstroms (Å) from the origin of the simulation box.

The average of IC, plotted against the cluster size as a function of radial position, is shown in fig. 2. The IC peaks at the center of the cluster and decreases rapidly upon approaching the liquid/cluster interface. The value of IC at the center of the smaller clusters is lower than the value at the center of larger clusters, consistent with the lower number density for small clusters shown in fig. 1. The highest IC value for the cluster center is most representative of the crystal. These observations are in line with results from DFT calculations [5] and the diffuse interface theory of nucleation [18-20] (see also chapter 4 in ref. [4]).

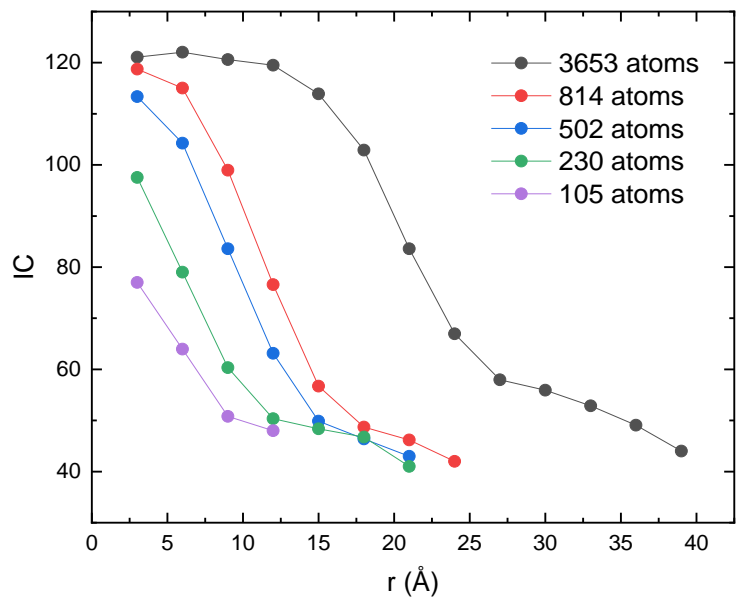

**Figure 2**- The value of the IC as a function of distance from the center of mass of five clusters of differing size at 1050K. The critical cluster size is 220, which is obtained in section 3.2.

3.2. Seeding

Studies of nucleation in a supercooled liquid require the value of the liquidus (or melting) temperature and the enthalpy of fusion. The coexistence method, involving both the crystal and liquid phase, was employed to determine the liquidus temperature, $T_m$, which was found to be 1525K. While there are no existing experimental results for this particular alloy, this liquidus temperature is in line with that of $Zr_{60}Ni_{21}Al_{19}$ ($T_m$ = 1423K [21]). The enthalpy of fusion ($h_m$), $2.05 \times 10^{-20}\ J/atom$, was obtained by calculating the difference in energy between the crystal and liquid phases at $T_m$. This was used to calculate the thermodynamic driving force for nucleation $\left(\Delta\mu = \dfrac{\Delta h_m \Delta T}{T_m}\right)$.

In fig. 3.a, a spherical cluster is inserted into the geometric center of the supercooled liquid. Before seeding, liquid atoms within this spherical region were removed to eliminate overlap. Multiple distinct seeds were introduced at varying temperatures. Detailed methodologies can be found elsewhere [22]. For illustration, figs. 3.b and 3.c show that a cluster at 1150K containing 371 atoms grew or dissolved in each simulation, indicating the inherent stochastic nature of nucleation. To enhance the reliability, each 30 simulations were made for each seed, with velocities randomly assigned before each run.

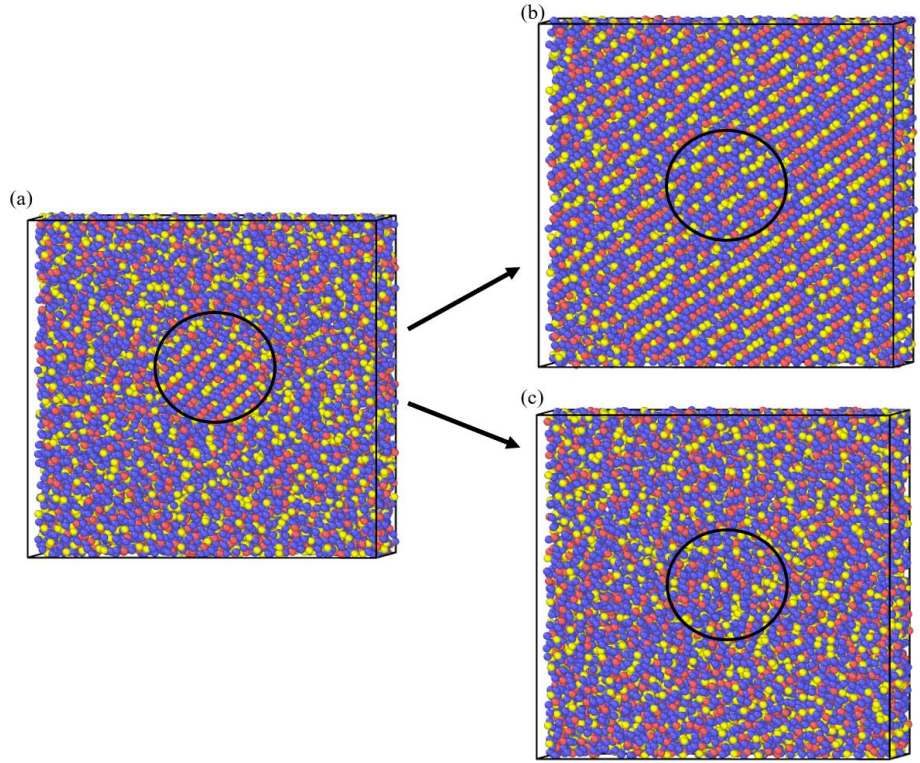

**Figure 3**- (a) 371 atoms inserted in the $Al_{20}Ni_{60}Zr_{20}$ metallic liquid at 1150K. (b) The cluster continued to grow after 1 ns; (c) the cluster completely dissolved after 1ns.

A linear fit was made to the percentage of the 30 simulations that grew as a function of the size of the seed. Since the probability for growth and dissolution are the same for the critical cluster size, the cluster size where 50% of the simulations grew is equal to $n^*$, as shown in fig. 4.a. Figure 4.b shows the calculated values of $n^*$ as a function of temperature from 1050K to 1250K. The number of atoms in the critical cluster increases from 220 to 1100 with increasing temperature, consistent with predictions from CNT. Below 1050K, $n^*$ is too small to effectively capture, resulting in a large error in determining its value. At temperatures above 1250K, $n^*$ is so large that it is too close to the boundary of the cell.

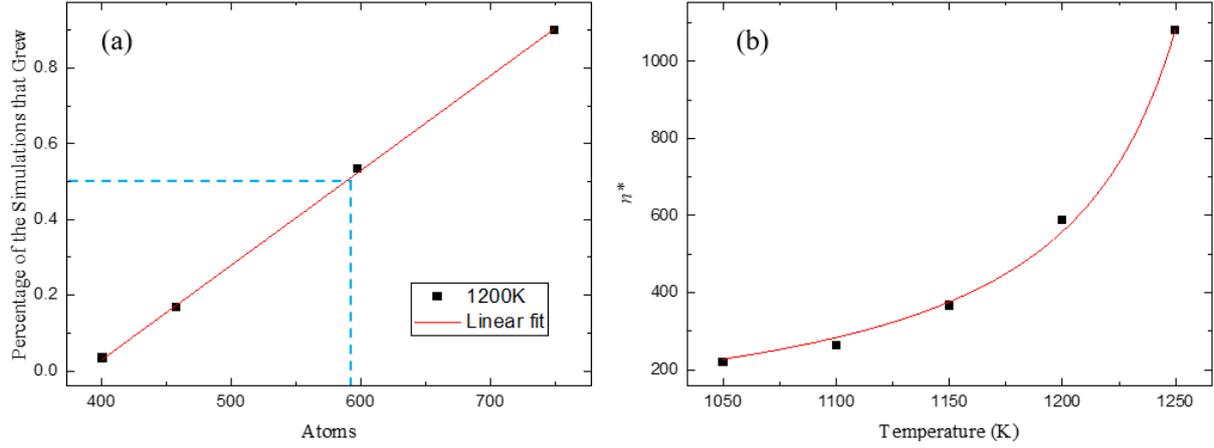

**Figure 4**- (a) The percentage of the 30 simulations that grew for the various cluster sizes at 1200K. The dashed line indicates the case where 50% of the simulations grew, corresponding to the critical size. (b) The critical cluster size, $n^*$, as a function of temperature; the red line is a fit to the CNT prediction.

The interfacial free energy, $\gamma$, was calculated from

$$n^* = \frac{32\pi\gamma^3}{3p^2|\Delta\mu|^3}, \qquad (2)$$

where $p$ [atom/m$^3$] is the atomic density of the crystalline cluster. The resulting value was essentially independent of temperature, with an average of 0.20 J/m². This is consistent with experimental measurements of alloys with related compositions and with MD studies on pure supercooled Ni, where $\gamma$ is approximately 0.22 J/m² [23]. The work of cluster formation at the critical cluster size scaled to $k_BT$ ($W^*/k_BT$) was obtained using $W^*$ calculated from the following equation

$$W^* = \frac{16\pi\gamma^3}{3p^2\Delta\mu^2}. \qquad (3)$$

The value of $W^*/k_BT$ increased from 44.6 at 1050K to 106.4 at 1250K. It is noteworthy that the magnitude of $W^*/k_BT$ is approximately two times larger than the typical value found in experimental studies [24, 25]. Combining the critical cluster size and the critical work of cluster formation, the Zeldovich factor, $Z^*$, can be computed

$$Z^* = \left( \frac{W^*}{3\pi (n^*)^2 k_B T} \right)^{\frac{1}{2}} . \qquad (4)$$

The value of $Z^*$ decreased from 0.01 at 1050K to 0.005 at 1200K. This range matches very well with existing MD and experimental studies [4, 22, 23]. Following Turnbull and Fisher, the attachment rate was estimated from the atomic jump rate from the liquid to the interface of the nucleating cluster, $6D/\lambda^2$, where $D$ is the diffusion coefficient in the liquid and $\lambda$ is the jump distance (assumed to be 3 Å). The self-diffusion coefficient in the $Al_{20}Ni_{60}Zr_{20}$ metallic liquid was calculated at the target temperature. Detailed information of the calculation on the diffusion can be found in previous work by our group [26].

Subsequently, the nucleation rate $I$ [$(m^3 s)^{-1}$] was obtained from

$$I = \frac{6D}{\lambda^2 v} Z^* \exp\left( -\frac{W^*}{k_B T} \right) , \qquad (5)$$

where $v$ is the molar volume of the liquid [$m^3$/atom]. The results are presented in Table 1. The liquid has a maximum nucleation rate at 1050K and rapidly decreases with increasing temperature. This trend is qualitatively consistent with predictions from CNT. The measured nucleation rates in similar alloys at the maximum reduced undercooling ($\Delta T = (T_m - T)/T_m$), which is about 0.2, are approximately $10^9$ m$^3$/s. At a reduced undercooling of 0.2 (corresponding to an undercooling temperature of 1220K), the predicted nucleation rate in $Al_{20}Ni_{60}Zr_{20}$ is also approximately $10^9$ m$^3$/s, supporting the accuracy of the MD calculations.

**Table 1**

Nucleation Parameters from the MD Simulations for Nucleation

| T (K) | $n^*$ (atom) | $\Delta\mu$ (J/atom) | $W^*/k_BT$ | $v$ (m$^3$/atom) | $Z^*$ | $D$ (m$^2$/s) | $I^{st}$ (m$^3$s)$^{-1}$ |
|---|---|---|---|---|---|---|---|
| 1050 | 220 | 6.39 × 10$^{-21}$ | 44.6 | 1.30 × 10$^{-29}$ | 0.010 | 2.4 × 10$^{-13}$ | 8.02 × 10$^{16}$ |
| 1100 | 263 | 5.71 × 10$^{-21}$ | 45.5 | 1.31 × 10$^{-29}$ | 0.009 | 3.6 × 10$^{-13}$ | 4.48 × 10$^{16}$ |
| 1150 | 366 | 5.04 × 10$^{-21}$ | 53.5 | 1.32 × 10$^{-29}$ | 0.007 | 5. × 10$^{-10}$ | 2.13 × 10$^{15}$ |
| 1200 | 589 | 4.37 × 10$^{-21}$ | 71.5 | 1.33 × 10$^{-29}$ | 0.005 | 7. × 10$^{-10}$ | 4.47 × 10$^{8}$ |
| 1250 | 1080 | 3.70 × 10$^{-21}$ | 106.4 | 1.33 × 10$^{-29}$ | 0.003 | 9. × 10$^{-10}$ | 3.68 × 10$^{-7}$ |

3.3. Mean First Passage Time

The MFPT is defined in a one-dimensional case as the average time that has elapsed until the system leaves a prescribed domain ($a$, $b$) around some initial point, $x_o$. The MFPT is calculated in terms of the time to go from $x_o$ to the final position, $b$, $\tau(x_o;a,b)$, which in general is given by [27]

$$\tau(x_o;a,b) = \int_{x_o}^{b} \frac{1}{D_o} \exp\left[\frac{U(y)}{k_BT}\right] dy \int_{a}^{y} \exp\left[-\frac{U(y)}{k_BT}\right] dz , \qquad (6)$$

where $D_o$ is the effective diffusion coefficient. To calculate the rate of transition, it is useful to calculate the time required for the system to reach the top of the energy barrier, $x^*$, $\tau(b=x^*)$. For nucleation, this would be the time to reach the critical size, $\tau(n^*)$. For that case, the rate at which the barrier is crossed, *i.e.* the nucleation rate, can be expressed in terms of this time

$$I = \frac{1}{2\tau(n^*)} . \qquad (7)$$

The factor of two arises because, at the top of the barrier, the system is equally likely to fall to either side of the barrier (as in the case of seeding at the critical size as discussed previously). Using MFPT, it is also possible to estimate the location of the transition state, $n^*$ in terms of the effective diffusion rate governing the attachment kinetics at the cluster interface, $D$,

$$\left.\frac{\partial^2 \tau(n)}{\partial n^2}\right|_{n=n^*} = \frac{1}{D} \quad . \tag{8}$$

Conversely, if the location of the transition state, $n^*$, is known, it is possible to determine the kinetic factor using MFPT. If the barrier is relatively high, the behavior of the MFPT near the critical size can be evaluated using the method of steepest descent [11] giving

$$\tau(n) = \frac{\tau_I}{2}\left(1 + erf\left(n-n^*\right)\varsigma\right) \tag{9}$$

where $erf$ is the error function, $\varsigma$ is the local curvature near the top of the barrier,

$$\varsigma = \sqrt{\frac{1}{2k_BT}\left.\frac{d^2W}{dn^2}\right|_{n=n^*}} \tag{10}$$

and $\tau_I = I^{-1}$, i.e., the inverse of the steady-state nucleation rate.

By evaluating the MFPT in simulation results and fitting to eq. (9) the nucleation rate, $I$, the critical size, $n^*$, and the curvature at the top of the barrier, which is related to the time lag in time-dependent nucleation, are obtained. This approach is widely used to analyze the results of MD simulations, particularly since small ensembles are sufficient [28]. For this study, a larger ensemble for the $Al_{20}Ni_{60}Zr_{20}$ liquid, containing 200,000 atoms, was prepared using the same procedure mentioned above. The sample was relaxed at the target temperatures for 2 ns to observe homogeneous nucleation. The size of the largest cluster in each simulation was recorded at regular intervals and the time at which each size appeared for the first time, $t_i(n)$, was also recorded. This time was averaged over ten repetitions to obtain the mean first-passage time. (This is illustrated in fig. 5.a for $\tau(100)$ at 1050K). The MFPT for each value of $n$ and $\tau(n)$ was then obtained by averaging over the values of $t_i(n)$. The values of $\tau(n)$ for different values of $n$ at 1050K were fit to eq. (10) in fig. 5.b to obtain the nucleation parameters, including the critical size.

For smaller cluster sizes, the increase in $\tau(n)$ exhibits a power-law dependence on cluster size, with the rate of increase slowing as the clusters grow larger. The MFPT fitting was extended to a larger dataset, specifically 500 data points instead of 200, as shown in fig. 5b. As a result, the critical cluster size grew from 91 to 106, and the nucleation rate decreased to $3.9\times 10^{33}(m^3s)^{-1}$. However, it was noted that the quality of the fit diminished, as evidenced by a lower $R^2$ value (0.99 to 0.97), when the data range was expanded.

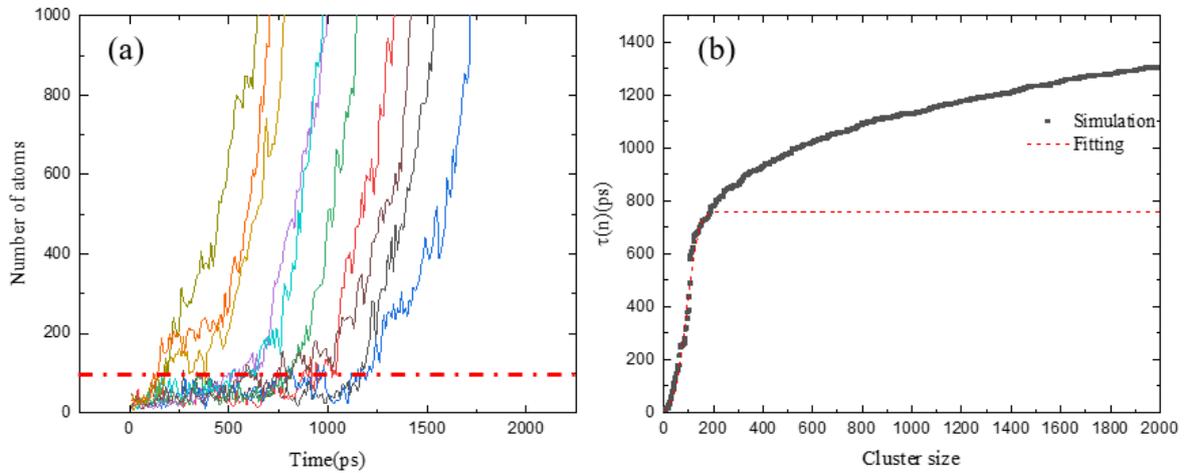

**Figure 5**- (a) The number of atoms in nucleating clusters as a function of time for ten different MD simulations at 1050K (each simulation is a different color line); $\tau(100)$ is shown for illustration by a red dot-dash line. (b) Fit of the different $\tau(n)$ values as a function of cluster size, $n$, to eq. (10) at 1050K.

The results are presented in Table 2. It is striking that the nucleation rate obtained is much higher than those normally observed in experimental studies of metallic liquids[4, 29, 30]. Further, it is seventeen orders of magnitude higher than the nucleation rate obtained from the seeding method at the same temperature (the seeding results were presented in the last section, 3.2). The critical cluster size deduced from MFPT is also roughly two to three times smaller than the value obtained from the seeding approach. Interestingly, the magnitude of the MFPT nucleation rates showcased in this study align with those observed for other metallic systems employing the MFPT technique [28, 31, 32]. When the critical size obtained by the MFPT at 1050K, $n^*$=91, was subjected to 30 iterations using the seeding method at the same temperature, every cluster swiftly dissolved within the initial 100 ps, indicating that this is not the critical size at that temperature. The pronounced discrepancy in predicted nucleation rates and critical sizes between the MFPT predictions, those obtained experimentally, and those obtained from the MD seeding method question the reliability of the MFPT method for nucleation studies.

**Table 2**

Nucleation Parameters from MFPT

| $T$ (K) | $n^*$ | $Z^*$ | $I^{st}$ (m$^3$s)$^{-1}$ |
|---|---|---|---|
| 900 | 70 | 0.0151 | 4.6× 10$^{33}$ |
| 1000 | 56 | 0.0148 | 6.8× 10$^{33}$ |
| 1050 | 91 | 0.0105 | 2× 10$^{33}$ |

## 3.4. Collective kinetics

As mentioned in the introduction, classical nucleation and growth models assume that the kinetics of crystallization are driven by single atom additions to the cluster [4]. However, recent MD studies of crystal nucleation in a liquid indicate that multiple liquid atoms attach to or detach from the cluster cooperatively [8]. These studies suggest that a group of neighboring atoms in the liquid next to the interface collectively make minor alterations in their order parameter to become incorporated into the nucleus. Over time, they adopt the same structural arrangement as the crystal cluster. For illustration, a target atom in the MD simulation that is identified as atom ID 29566, was randomly selected. Figures 6.a and 6.b show the IC value and the local potential energy (PE) of this atom as a function of time. A sudden increase in the IC value within the time range from 410 ps to 470 ps indicates attachment of that atom to the cluster.

Also shown in fig. 6.a and 6.b are the IC and local PE values of the target's neighboring atoms. Neighboring atoms are defined as those atoms that remained within 3.5 Å of the target atom for more than 80% of the observed time. Their change in IC as a function of time mirror that of the target atom. A dispersion in the local PE values is noted before 410 ps, but they quickly stabilize upon atom attachment. The starting point of the rapid stabilization corresponds well with that of the IC changes. Such synchronized behavior amongst neighboring atoms underscores the notion that atoms collectively attach to the cluster.

A similar investigation of the IC and local PE energy was made for a dissolving cluster. In fig. 6.c, a target atom (atom ID 22298) and its neighboring atoms exhibit a decreasing IC value with time, eventually detaching from the cluster at 120 ps. After detachment their local PE becomes more unstable, showing a trend opposite to that observed in the attachment case. Nucleation is a stochastic process, with small clusters both growing and shrinking. Figure 6 demonstrates that the IC values and the local PEs of both the target atom and its neighboring atoms act cooperatively during attachment and detachment.

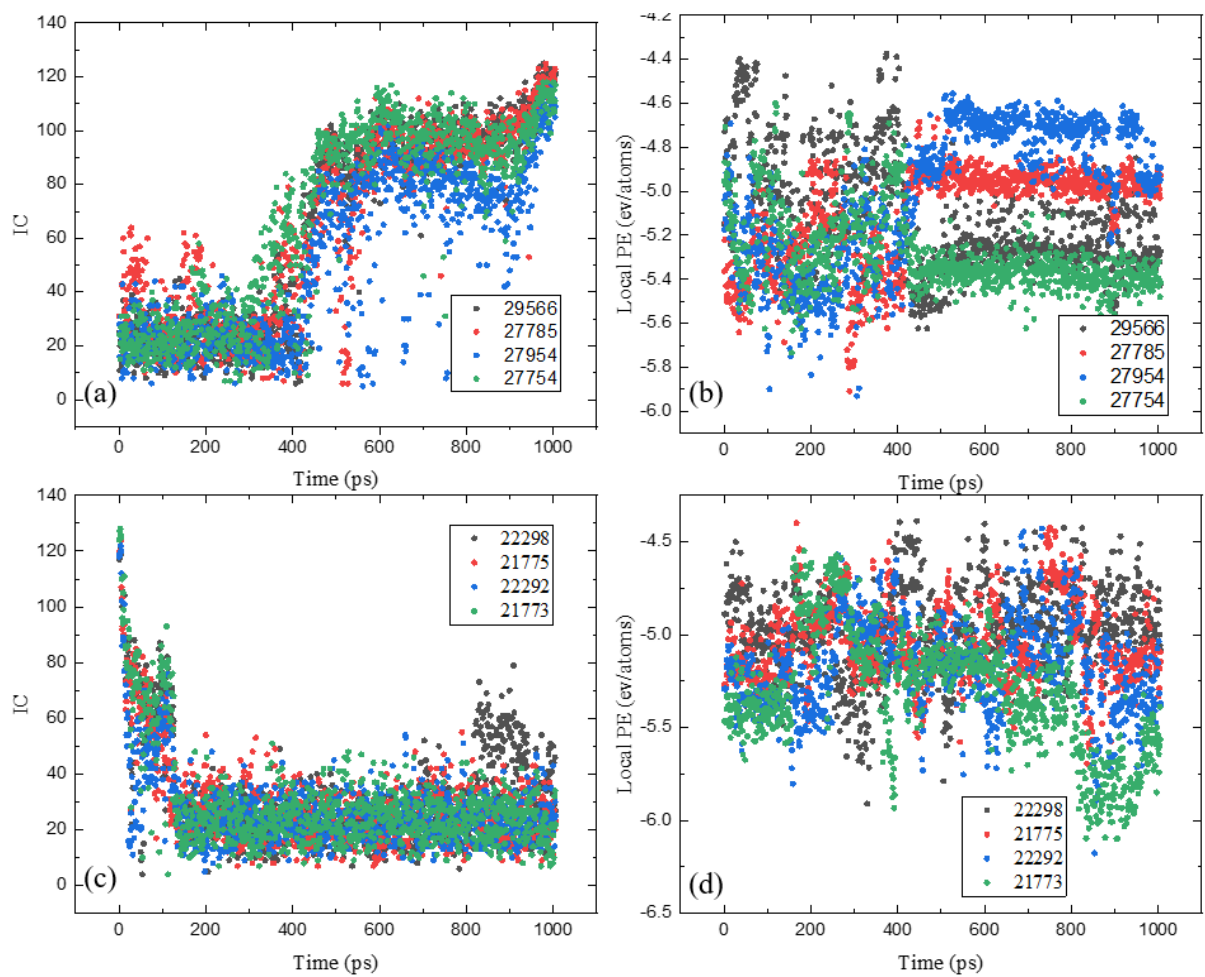

**Figure 6**- (a) IC and (b) local potential energy (PE) as a function of time for atom attachment. (c) and (d) represents the IC change and corresponding PE as a function of time for atom detachment.

3.5. Collective motion in nuclei coalescence

The classical nucleation pathway assumes that nucleation and growth are achieved via atom – atom addition (or monomer–monomer addition). However, several non-classical nucleation pathways have been proposed and found, including coalescence, Ostwald ripening, oriented attachment, and others. Due to the rapid development of the in-situ observations, the coalescence of growing clusters into single crystals or crystals with grain boundaries has been experimentally observed [9, 10]. However, the atomistic understanding of such process remains unclear. An intriguing question arises: Does cooperativity exist during the coalescence?

To mimic this process, two clusters were placed 4 Å apart and integrated into the $Al_{20}Ni_{60}Zr_{20}$ metallic liquid for an MD simulation. Two cases were investigated. In one case the two clusters had the same orientation. In the second case, one of the clusters was oriented at a 45-degree angle along the x-axis with respect to the other cluster. The simulation was made at 1050K, with each cluster containing 177 atoms, which is smaller than the critical cluster size ($n^* = 220$) at this temperature.

The sample was relaxed for 1 ns, during which nucleation was observed. For the case where the two clusters had the same orientation, they rapidly merged and showed continued growth, as shown in fig. 7.a. A target atom was randomly selected from the merged area, and the neighboring atoms to that atom were identified. Their IC values are presented as a function of time in fig. 7.b. The rapid increase of IC values aligns well with the observed rapid merging of the two clusters. In contrast, the misoriented clusters eventually formed a grain boundary after 500 ps, as shown in fig. 7.c. Instead of the sudden increase in IC observed for clusters with the same orientation, the IC value changed more gradually for the misoriented clusters (fig. 7.d.). The color represents the IC level, with red representing the largest value of IC and blue the lowest. For better visualization, only atoms with an IC value greater than 50 are represented. The interior atoms exhibit the highest IC values, while the surface atoms have the lowest. The phenomena of connected clusters and grain boundary formation has been observed in recent in-situ growth experiments in amorphous bismuth [9]. These MD results show that the same can happen during nucleation.

To understand the cooperative behavior of atoms during coalescence, particularly those located at the interface between the two clusters, a similar procedure to that described earlier to identify neighboring atoms and calculate the average coherence length was followed. By randomly selecting at least 100 atoms from inside and outside the interface region, the coherence length was found to be approximately 10 atoms at 1050K in both scenarios. This coherence number agrees with our previous studies of attachment and detachment during nucleation from the liquid [8]. To study the cooperative behavior for atoms located at different regions, the average time required for the atoms within and without the interface to change IC from 40 to 80 was investigated. The average time indicates the speed of the cooperative atoms transforming from liquid atoms to solid atoms. For clusters having the same orientation, the time for the interface atoms to change was approximately 1.5 times faster than for the atoms outside the interface, coinciding the rapid coalescence. For the misaligned clusters, the time for the interface atoms was between 1.5 to 2 times longer than for the atoms outside the interface. This observation aligns with visual representations in figs. 7.a and 7.c, which show that clusters either rapidly coalesce or eventually develop a grain boundary. An extension of the studies to clusters of varying sizes and orientations showed that the cooperative motion was consistently present across all examined scenarios.

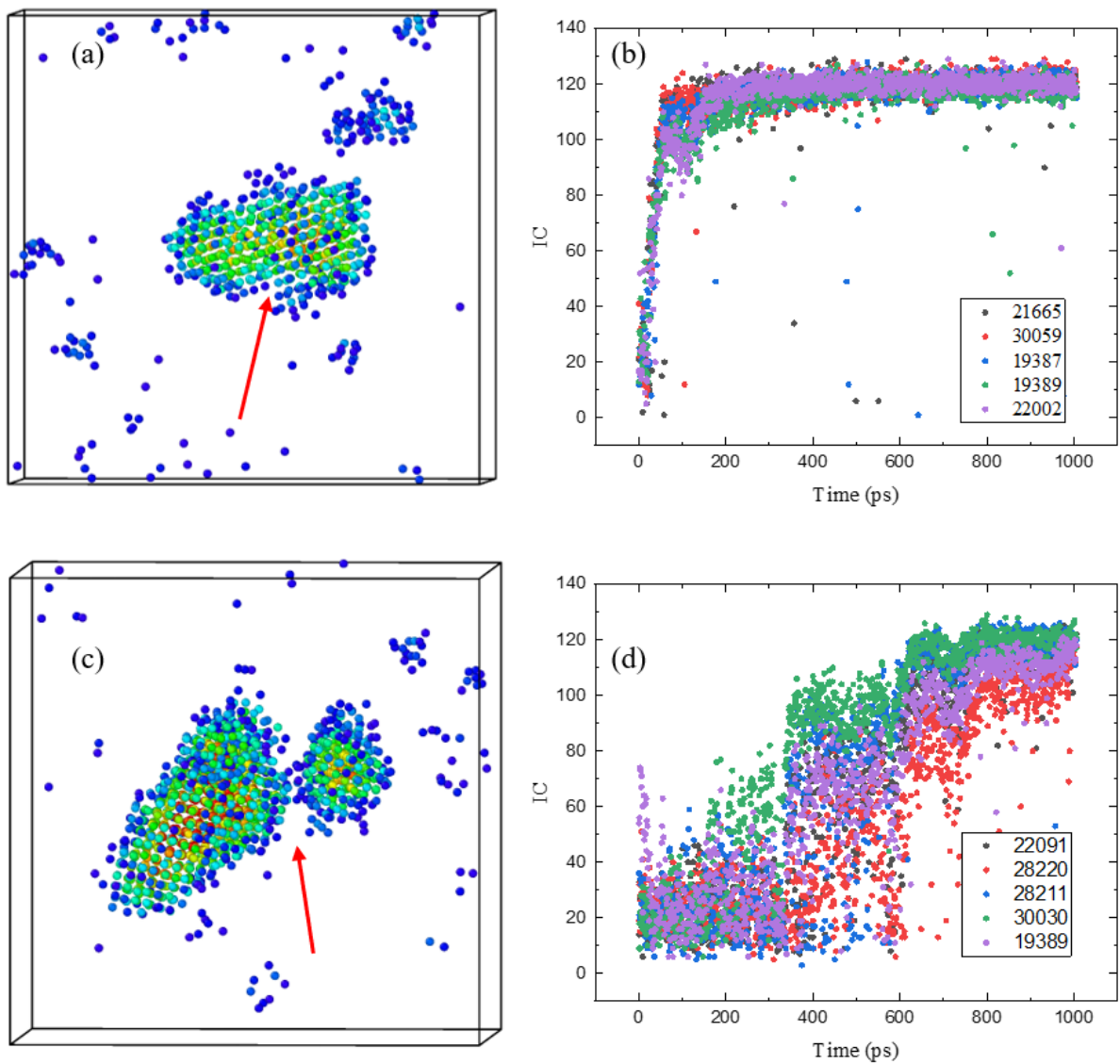

Figure 7 – (a) Two clusters with same orientation connected in the early stage of the simulation. (b) The IC value, plotted as a function of time, corresponds to a randomly selected atom and its neighbor atoms from the interface between the two clusters shown in (a). The IC quickly increased in the first 100 ps. (c) A clear boundary was found for the mis-oriented clusters after 500 ps. (d) The IC value as a function of time from the target atom and its neighbor atoms randomly selected from the boundary; the increase is more gradual.

## 4. Conclusion

Crystal nucleation in a $Al_{20}Ni_{60}Zr_{20}$ metallic liquid was studied using MD simulations. The results show that the nucleating cluster is neither spherical nor compact, a finding consistent with prior research . The critical cluster size was determined within the context of CNT and the calculated nucleation rates agreed reasonably well with existing experimental data and with these MD results. This supports the validity of CNT for the study of metallic liquids. However, discrepancies arose when comparing the critical cluster size and nucleation rate derived from the MFPT method with those obtained from the seeding method. Specifically, with a considerably smaller critical cluster size (90 as opposed to 220), the MFPT nucleation rate was nearly 17 orders of magnitude greater than the CNT nucleation rate. This vast difference indicates that the use of the MFPT method requires caution. Contrary to the individual diffusive jumps assumed by CNT, the MD simulations showed that nucleation is achieved via collective motion. Our results illustrate that the kinetics of cooperative attachment and detachment involve neighboring atoms acting simultaneously during nucleation. The local potential energy of these neighboring atoms concurrently stabilizes or destabilizes during the attachment or detachment. Beyond the classical nucleation pathway, we observed atom cooperativity in non-classical nucleation processes, specifically during the coalescence of nuclei. This suggests that collective motion might be a pervasive phenomenon across various nucleation processes. These insights into atomic-scale motions contribute significantly to the understanding of nucleation and the following growth processes and suggest that collective behavior should be incorporated in future nonclassical theories of nucleation and growth.

**Acknowledgements**

We thank Anupriya Agrawal and Ryan Chang for scientific discussions and preliminary studies that inspired the studies presented here. The research was partially supported by the National Science Foundation under Grant No. DMR-19-04281. Any opinions, findings, and conclusions or recommendations expressed in this material are those of the author(s) and do not necessarily reflect the views of the National Science Foundation.